\documentclass[fleqn,10pt]{wlscirep}
\usepackage[utf8]{inputenc}
\usepackage[T1]{fontenc}
\usepackage{lineno}
\usepackage{bm}
\usepackage{multirow}
\usepackage{makecell}
\usepackage{amsmath, amsthm, amssymb, amsfonts}

\usepackage{latexsym}
\def\hbar{\mathit{^^^^0127}}

\usepackage{color,soul}

\title{QD3SET-1: A Database with Quantum Dissipative Dynamics Data Sets}

\author[1]{Arif Ullah}
\author[2]{Luis E. Herrera Rodr\'iguez}
\author[1,*]{Pavlo O. Dral}
\author[2,*]{Alexei A. Kananenka}
\affil[1]{State Key Laboratory of Physical Chemistry of Solid Surfaces, Fujian Provincial Key Laboratory of Theoretical and Computational Chemistry, Department of Chemistry, and College of Chemistry and Chemical Engineering, Xiamen University, Xiamen 361005, China}
\affil[2]{Department of Physics and Astronomy, University of Delaware, Newark, Delaware 19716, United States}

\affil[*]{corresponding author(s): Pavlo O. Dral (dral@xmu.edu.cn) and 
Alexei A. Kananenka (akanane@udel.edu)}

\begin{abstract}
Simulations of the dynamics of dissipative quantum systems
utilize many methods such as physics-based quantum, semiclassical, and quantum-classical as well as machine learning-based approximations, development and testing of which requires diverse data sets. 
Here we present a new database QD3SET-1 containing eight data sets of quantum dynamical data for two systems of broad interest, spin-boson (SB) model and the Fenna--Matthews--Olson (FMO) complex, generated with two different methods solving the dynamics, approximate local thermalizing Lindblad master equation (LTLME) and highly accurate hierarchy equations of motion (HEOM).
One data set was generated with the SB model which is a two-level 
quantum system coupled to a harmonic environment using HEOM for 1,000 model parameters. 
Seven data sets were collected for the FMO complex of different sizes
(7- and 8-site monomer and 24-site trimer with LTLME and 8-site monomer with HEOM) for 500--879 model parameters. Our QD3SET-1 database contains both population and coherence dynamics data and part of it has been already used for machine learning-based quantum dynamics studies.

\end{abstract}
\begin{document}

\flushbottom
\maketitle

\thispagestyle{empty}


\section*{Background \& Summary}
The simulation of the inherently quantum-mechanical dynamics underlying charge,
energy, and coherence transfer in the condensed-phase is one of the most difficult 
challenges for computational physics and chemistry. 
The exponential scaling of the computational cost with system size makes 
the quantum-mechanically exact simulations of such processes in complex systems infeasible. 
With the exception of a few model Hamiltonians whose form makes the
numerically exact quantum-dynamics simulations possible, any simulation of general
condensed-phase systems must rely on 
approximations.\cite{meyer09,makri99,meyer90,wang03,tanimura20,tanimura89a,tanimura90,greene17,kapral06,kapral16,min17,min15,gao20,crespootero18,subotnik16,wang16,mclachlan06,tully90,shushkov12,huo13,kapral99,miller16,sun16,chenu15,han2020stochastic,ullah2020stochastic,yan2022piecewise,chen2022universal} Data-driven machine learning (ML) methods for quantum dynamics emerged as attractive alternative to the physics-based
approximations due to their low computational cost and high accuracy.~\cite{rodriguez21,rodriguez22,ullah2022one,ullah2022predicting,ullah2021speeding,naicker22,akimov21,secor21,banchi18,bandyopadhyay18,yang20,wu21,lin21,tang22,lin2022realization,lin22,choi22,zhang2023excited} Development and testing of new simulation methodologies, both physics- and ML-based, would be greatly facilitated if high-quality reference quantum-dynamics data for a diverse set of quantum systems of interest were available.

Here we present a QD3SET-1 database, a collection of eight data sets of time-evolved 
population dynamics of the two
systems: spin-boson (SB) model and the Fenna--Matthews--Olson (FMO) light-harvesting complex. The data sets are summarized in Table~\ref{tab:tab1}.
The SB model describes a (truncated or intrinsic)
two-level quantum system linearly coupled to a harmonic bath.\cite{leggett87,weiss12} 
The physics of both the ground state and the dynamics of the SB model is very 
rich. It has been a continuous object of study during the past decades.
SB has become a paradigmatic model system in the development of approximate
quantum-dynamics methods 
and, nowadays, it is becoming a popular choice for the development of ML models.\cite{ullah2021speeding,wu21,rodriguez22}

The FMO system 
has become one of the most extensively studied natural light-harvesting complexes.\cite{adolphs06,ishizaki09,panitchayangkoon10,zerah21,ritschel11,shim12,fenna75}
Under physiological conditions, the FMO complex forms a homotrimer consisting of eight bacteriochlorophyll-a (BChla)
molecules per monomer. 
The biological function of the FMO trimer is to transfer excitation energy from the 
chlorosome to the 
reaction center (RC).\cite{milder10} 
An interest in this light-harvesting system sparked when two-dimensional electronic spectroscopy 
experiments detected the presence of quantum coherence effects in the FMO complex.~\cite{engel07,scholes17,engel11} These observations
triggered intense debates about the role this coherence might play in the highly efficient excitation energy transfer (EET).

Early studies of the FMO complex considered only seven-site FMO models comprising of BChla 1--7. 
Until BChla 8 was discovered, BChla 1 and 6 were both 
assumed to be possible locations for accepting the excitation from the chlorosome
because they are believed to be the nearest pigments to the antenna which captures the sunlight.\cite{renger98,louwe97,list13}
From there, the energy is subsequently funneled through two nearly independent routes:
from site 1 to 2 (pathway 1) or from site 6 to sites 7, 5, and 4 
(pathway 2). 
The terminal point of either route is site 3, where the exciton is then transferred to the RC.\cite{moix11}

Ever since the discovery of the eighth BChla, the role of this pigment in the EET has 
been extensively investigated.\cite{busch11,huang12,bina13,olbrich11,ritschel11,muhlbacher12,tronrud09,moix11,jia15} 
In particular, it was shown that while the population dynamics of the eight-site FMO model is markedly different 
from a seven-site configuration, the EET efficiencies in both models  were predicted to
remain comparable and very high.\cite{moix11} BChla 8 has also been suggested as
possible recipient of the initial excitation.

The dynamics of the FMO model has been a subject of numerous computational studies primarily focusing on understanding 
the role of the protein environment on
the efficiency of EET (see e.g., Refs.~\citenum{shabani12,wu10,suzuki20,mohseni14}).
Numerical simulations typically employ one of the several parameterized or fitted into the experimental data
FMO model Hamiltonians that differ in the BChla excitation energies and the couplings between different BChla sites.\cite{milder10,vulto98,adolphs06,cho05,hayes11,kell16,rolczynski21} Simulations of the full FMO trimer containing
 24 BChl have also been performed.\cite{ke16,wilkins15} 

Reported in this Data Descriptor a QD3SET-1 database contains 
seven data sets of time-evolved population dynamics of FMO models with different system Hamiltonians and initial excitations
for several hundreds of bath and system-bath parameters. 
Hierarchy of equations of motion (HEOM) approach\cite{tanimura90,tanimura20} was used to simulate the population dynamics of SB and FMO models,
in one of the seven FMO data sets. HEOM is a numerically exact
method that can describe the dynamics of a system with a non-perturbative and non-Markovian system–bath interaction. The high computational
cost of HEOM, however, limits the number of FMO simulations that can be performed with this method. To generate other six FMO
data sets, 
an approximate method---local thermalizing Lindblad master equation (LTLME)\cite{bourne2019structure, joseph2020thesis} was used.

Some of our data was already used in previous studies developing and benchmarking ML models for quantum-dynamics simulations.~\cite{ullah2021speeding,ullah2022predicting,rodriguez22,ullah2022one} Here we regenerate one of the data sets to augment with more data and provide many new data sets generated from scratch (Table~\ref{tab:tab1}). To facilitate their use, we organize the data sets in a coherently formatted database and provided metadata and extraction scripts.
We expect our Database that accompany this Data Descriptor 
will serve as valuable resources in the development of new quantum-dynamics methods.

\section*{Methods}
\subsection*{SB data set}
This data set is re-generated with the same settings and the same parameters as in on our previous SB data set\cite{ullah2021speeding} in order to include all the elements (populations and coherences) of the system's reduced density matrix (RDM).
The populations and population differences were published and used before.\cite{ullah2021speeding,rodriguez22}
Below we provide a brief summary for self-containing presentation of the data set.

\subsubsection*{Spin-boson model}
The spin-boson model
comprises a two-level quantum subsystem (TLS) coupled to a bath of harmonic oscillators. The
Hamiltonian has the following standard system-bath form: $\hat{H}=\hat{H}_s+\hat{H}_b+\hat{H}_{sb}$.
The Hamiltonian of the TLS in the  local (or site) basis $\{|+\rangle, |-\rangle\}$ is given by ($\hslash$=1)
\begin{equation}
    \hat{H}_s = \epsilon\left( |+\rangle\langle +| - |-\rangle\langle -|\right) \, + \, \Delta\left( |+\rangle\langle -| + |-\rangle\langle +|\right),
\end{equation}
where  $\epsilon$ is the so-called energetic bias and $\Delta$ is the tunneling splitting. The harmonic bath is an ensemble
of independent harmonic oscillators
\begin{equation}
    \hat{H}_b = \sum_{j=1}^{N_{b}} \left( \frac{\hat{p}_j^2}{2m_j} + \frac{1}{2}m_j\omega_j^2 \hat{x}_j^2  \right), \label{eq:hb}
\end{equation}
where $\{\hat{x}_j\}$ and $\{\hat{p}_j\}$ are the coordinates and momenta, respectively, of $N_b$ independent
harmonic bath modes with masses $\{m_j\}$ and frequencies $\{\omega_j\}$.
The TLS and bath are coupled through the additional term
\begin{equation}
    \hat{H}_{sb} = -\sum_{j=1}^{N_b} c_j \hat{x}_j\left( |+\rangle\langle+| -|-\rangle\langle -|\right) ,
\end{equation}
where $\{c_j\}$ are the coupling coefficients.

The  effects  of  the  bath  on the dynamics of TLS are collectively  determined  by the spectral density
function\cite{caldeira83}
\begin{equation}
J(\omega)=\frac{\pi}{2} \sum_{j=1}^{N_b} \frac{c_j^2}{m_j\omega_j}\delta(\omega - \omega_j). \label{eq:sd}
\end{equation}
In this work we choose to employ the Debye form of the spectral density (Ohmic spectral 
density with the Drude--Lorentz cut-off)~\cite{wang99}
\begin{equation}
    J(\omega)=2\lambda \frac{\omega \gamma}{\omega^2 + \gamma^2}, \label{eq:dl}
\end{equation}
where $\lambda$ is the bath reorganization energy, which controls the strength of system-bath coupling, 
and the cutoff frequency $\gamma=1/\tau_c$ ($\tau_c$ is the bath relaxation time).

All dynamical properties of the TLS can be obtained from the RDM
\begin{equation}
\tilde{\rho}_{\alpha \beta}(t) = \text{Tr}_b \langle \alpha| e^{-i\hat{H}t/\hslash}\hat{\rho}(0)e^{i\hat{H}t/\hslash}|\beta\rangle,  
\end{equation}
where $\alpha,\beta \in \{|+\rangle,|-\rangle\}$, 
$\hat{\rho}$ is the total density operator, and 
the trace is taken over bath degrees of freedom. 
For example, the commonly used in benchmark studies 
population difference is obtained from the RDM as follows: 
$p_+(t)-p_-(t) = \tilde{\rho}_{++}(t) -\tilde{\rho}_{--}(t) $.

The initial state of the total system is assumed to be a product state of the system and bath
 in the following form
\begin{equation}
\hat{\rho}(0) = \hat{\rho}_\mathrm{s}(0)\hat{\rho}_\mathrm{b}(0). \label{eq:rho0}
\end{equation}
In Eq.~\eqref{eq:rho0} the bath density operator is an equilibrium canonical density operator 
$\hat{\rho}_b(0) = e^{-\beta \hat{H}_\mathrm{b}}/\mathrm{Tr}_\mathrm{b}\left[e^{-\beta \hat{H}_\mathrm{b}}\right]$, 
where $\beta=(k_\mathrm{B}T)^{-1}$ is the inverse temperature and $k_\mathrm{B}$ is the
Boltzmann constant. The initial density operator of the system is chosen to be $\hat{\rho}_\mathrm{s}(0)=|+\rangle \langle +|$.
These conditions corresponds to instantaneous photoexcitation of the subsystem.

\subsubsection*{Data generation with spin-boson model and the hierarchy equations of motion approach} 
The data set for the spin-boson model was generated as described previously.\cite{ullah2021speeding} We also summarize it  below. 
The following system and bath parameters were chosen:
 $\tilde{\epsilon}=\epsilon/\Delta=\{0, 1\}$, $\tilde{\lambda}=\lambda/\Delta=\{0.1, 0.2, 0.3, 0.4, 0.5, 0.6,0.7, 0.8, 0.9, 1.0\}$, 
$\tilde{\gamma}=\gamma/\Delta =\{1, 2, 3, 4, 5, 6, 7, 8, 9, 10\}$, and $\tilde{\beta}=\beta \Delta=\{0.1, 0.25, 0.5, 0.75, 1\}$,
where the tunneling matrix element $\Delta$ is set as an energy unit. 
For all combinations of these
parameters the system's RDM was propagated using HEOM
approach 
implemented in \textsc{QuTiP} software package.~\cite{johansson12} 
The total propagation time was $t_\mathrm{max} \Delta=20$ and the HEOM integration time-step was set to $dt \Delta=0.05$. 
In total, 1,000 of HEOM calculations, 500 for symmetric ($\epsilon/\Delta=0$) and 500 for asymmetric ($\epsilon/\Delta=1$)
spin-boson Hamiltonian 
were performed. The data set contains a set of RDMs 
from $t \Delta=0$ to $t_\mathrm{max} \Delta=20$, saved every $dt\Delta=0.05$,
for every combination of the parameters $(\tilde{\epsilon},\tilde{\lambda},\tilde{\gamma},\tilde{\beta})$ 
described above. 

\subsection*{Fenna--Matthews--Olson complex data sets}
In this section we first describe the general theory behind the FMO model Hamiltonian and later for each data set we provide specific technical details. See also Table~\ref{tab:tab1} for an overview of each data set.
\subsubsection*{FMO model Hamiltonian}
The FMO complex in this work is 
described by the system-bath Hamiltonian with the renormalization term
$\hat{H} = \hat{H}_s + \hat{H}_b + \hat{H}_{sb} + \hat{H}_{ren}$. The electronic system
 is described by the Frenkel exciton Hamiltonian
\begin{equation}
    \hat{H}_s = \sum_{n=1}^{N_e} E_n|n\rangle \langle n| + \sum_{n,m=1,n \neq m}^{N_e} V_{nm} |n \rangle \langle m|, \label{eq:hs}
\end{equation}
where $|n\rangle$ denotes that only the $n$th site is in its electronically excited state
and all other sites are in their electronically ground states, $E_n$ are the transition energies, 
and $V_{nm}$ is the Coulomb coupling
between $n$th and $m$th sites. The couplings are assumed to be constant 
(the Condon approximation).
Note that the overall electronic ground state
of the pigment protein complex $|0\rangle$ is assumed to be only radiatively coupled to the single-excitation manifold
and as such it is not included in the dynamics calculations.
In analogy with the SB model the bath is modeled by a set of independent harmonic oscillators. The thermal bath is coupled 
to the subsystem's states $|n\rangle$ through the system-bath interaction term
\begin{equation}
    \hat{H}_{sb} = \sum_{n=1}^{N_{e}} \sum_{j=1}^{N_{b}}  c_{nj} \hat{x}_j |n\rangle \langle n|, \label{eq:hsb}
\end{equation}
where each subsystem's state is independently coupled to its own harmonic environment and $c_{nj}$ are the
pigment-phonon coupling  constants of environmental phonons local to the $n$th BChla.

The FMO model Hamiltonian contains a reorganization term which counters the shift in the minimum energy positions of
harmonic oscillators introduced by the system-bath coupling. In the case that each state $|n\rangle$ is independently coupled
to the environment the renormalization term takes the following form
\begin{equation}
    \hat{H}_{ren} = \sum_{n=1}^{N_{e}} \lambda_n |n\rangle \langle n|,
\end{equation}
where $\lambda_n = \sum_j c_{nj}^2/(2m_j\omega_j^2)$ is the bath reorganization energy. 
The bath spectral density associated with each electronic state is assumed to be given by the Lorentz--Drude 
spectral density (Eq.~\ref{eq:dl}).

Analogously to the SB data set the initial state of the total system is assumed to 
be a product state of the system and bath. The initial electronic density operator given by
$\hat{\rho}_s(0)$ was varied as described below. 
The bath density operator is taken to be the equilibrium canonical density operator.

\subsubsection*{FMO-Ia, FMO-Ib, and FMO-II data sets: 7-site FMO models with the local thermalizing Lindblad master equation approach}
We generated data sets for 
 the two 7-site system ($N_{e}=7$) Hamiltonians. FMO-I data set was generated for the system Hamiltonian  parameterized by Adolphs and Renger\cite{adolphs06} and given by (in cm$^{-1}$)
\begin{equation}
    H_s = \begin{pmatrix}
    200 & -87.7 & 5.5 & -5.9 & 6.7 & -13.7 & -9.9 \\
    -87.7 & 320 & 30.8 & 8.2 & 0.7 & 11.8 & 4.3 \\
    5.5 & 30.8 & 0 & -53.5 & -2.2 & -9.6 & 6.0 \\
    -5.9 & 8.2 & -53.5 & 110 & -70.7 & -17.0 & -63.6 \\
    6.7 & 0.7 & -2.2 & -70.7 & 270 & 81.1 & -1.3\\
    -13.7 & 11.8 & -9.6 & -17.0 & 81.1 & 420 & 39.7 \\
    -9.9 & 4.3 & 6.0 & -63.3 & -1.3 & 39.7 & 230 
    \end{pmatrix}, \label{eq:fmo1-1}
\end{equation}
FMO-Ia data set comes directly from our previous studies\cite{ullah2022predicting,ullah2022one} and FMO-Ib data set
was generated here for a broader parameter space as described below. 

FMO-II data set was generated for the Hamiltonian parameterized by Cho \textit{et al.}\cite{cho05} which takes the following form (in cm$^{-1}$)
\begin{equation}
    H_s = \begin{pmatrix}
280 & -106 & 8 & -5 & 6 & -8 & -4 \\
-106 & 420 & 28 & 6 & 2 & 13 & 1 \\
8 & 28 & 0 & -62 & -1  & -9 & 17 \\
-5 & 6 & -62 & 175 & -70 & -19 & -57 \\
6 & 2 & -1 & -70 & 320 & 40 & -2 \\
-8 & 13 & -9 & -19 & 40 & 360 & 32 \\
-4 & 1 & 17 & -57 & -2 & 32 & 260 
\end{pmatrix}. \label{eq:fmo1-2}
\end{equation}
The diagonal 
offset of 12210 cm$^{-1}$ is added to both Hamiltonians. 
Each site is coupled to its own bath characterized by the Drude--Lorentz spectral density, Eq.~\ref{eq:dl}, but the bath of each site
is described by the same spectral
density. 

For FMO-Ia data set, the following spectral density parameters and temperatures were 
employed: $\lambda$ = $\{$10, 40, 70, \ldots, 310$\}$ cm$^{-1}$,
$\gamma$ = $\{$25, 50, 75, \ldots, 300$\}$ fs rad$^{-1}$,
and T = $\{$30, 50, 70, \ldots, 310$\}$ K.
For FMO-Ib and FMO-II data sets, the spectral density parameters and temperatures were: $\lambda$ = $\{$10, 40, 70, \ldots, 520$\}$ cm$^{-1}$,
$\gamma$ = $\{$25, 50, 75, \ldots, 500$\}$ cm$^{-1}$,
and T = $\{$30, 50, 70, \ldots, 510$\}$ K. 

For FMO-Ia, FMO-Ib, and FMO-II data sets, the farthest-point sampling\cite{dral2019mlatom} was employed to select the most distant points in the Euclidean space\cite{ullah2022predicting} of parameters which typically more efficiently covers relevant space compared to random sampling.\cite{dral2019mlatom} 
We choose the top 500 (most distant)
combinations of ($\lambda$ , $\gamma$, $T$) based on farthest-point sampling.
 For each selected set of parameters the system RDM was calculated
using the local thermalizing Lindblad master equation (LTLME) approach\cite{bourne2019structure, joseph2020thesis} 
implemented in the \textsc{quantum\_HEOM} package.\cite{joseph2019quant,joseph2020thesis}
Two subsets of the data set were generated, one for the 
initial electronic density operator $\hat{\rho}_s(0)=|1\rangle\langle 1|$ corresponding to the initial excitation of 
 site-1 and the other one for the initial density operator
 $\hat{\rho}_s(0)=|6\rangle\langle 6|$  which corresponds to the initial excitation of site-6. In each case,
500 RDM trajectories were generated.  The data set contains both diagonal (populations)
and off-diagonal (coherences) elements of the RDM on a time grid from 0 to 1~ns (in the case of FMO-Ia) and 0 to 50~ps (in the case of FMO-Ib and FMO-II) with the 5 fs time step. 

\subsubsection*{FMO-III and FMO-IV data sets: 8-site FMO models with the local thermalizing Lindblad master equation approach}
Using the same LTLME-based approach, we generated a data set for two different Hamiltonians for the 
8-site FMO model. The first Hamiltonian (FMO-III data set) was parameterized by
Jia \textit{et al.}\cite{jia2015hybrid} 
The electronic system Hamiltonian is given by (in cm$^{-1}$) 
\begin{equation}
    H_s = \begin{pmatrix}
218 & -91.0 &  4.1 & -6.3 & 6.3 & -8.8 & -7.8 & 32.4 \\
-91.0 & 81 & 28.7 &  8.2  & 1.0 & 8.8 & 3.4 & 6.3 \\
 4.1 & 28.7 & 0 & -46.6 & -4.4 & -9.3 & 1.3 & 1.3 \\
-6.3 & 8.2 & -46.6 & 105 & -73.9 & -17.7 & -59.1 & -1.9 \\
 6.3 & 1.0 & -4.4 & -73.9 & 105 & 76.0 & -3.1 & 4.2 \\
-8.8 & 8.8 & -9.3 & -17.7 & 76.0  & 186 & 25.9 & -11.6 \\
-7.8 & 3.4 & 1.3 & -59.1 & -3.1 & 25.9 & 169 & -11.9 \\
32.4 & 6.3 &  1.3 & -1.9 & 4.2 & -11.6 & -11.9 & 154
\end{pmatrix}, \label{eq:fmo2-2}
\end{equation}  
with the diagonal offset of 11332 cm$^{-1}$.   

The FMO-IV data set was generated for the Hamiltonian parameterized by 
Busch \textit{et al.}\cite{busch11} (site energies) and Olbrich \textit{et al.}~\cite{olbrich11} (excitonic couplings) and takes the
following form (in cm$^{-1}$) 
\begin{equation}
    H_s = \begin{pmatrix}
310 & -80.3 & 3.5 & -4.0 & 4.5 & -10.2 & -4.9 & 21.0\\
-80.3 & 230 & 23.5 & 6.7 & 0.5 & 7.5 & 1.5 & 3.3\\
3.5 & 23.5 & 0 & -49.8 & -1.5 & -6.5 & 1.2 & 0.7\\
-4.0 & 6.7 & -49.8 & 180 & 63.4 & -13.3 & -42.2 & -1.2\\
4.5 & 0.5 & -1.5 & 63.4 & 450 & 55.8 & 4.7 & 2.8\\
-10.2 & 7.5 & -6.5 & -13.3 & 55.8 & 320 & 33.0 & -7.3\\
-4.9 & 1.5 & 1.2 & -42.2 & 4.7 & 33.0 & 270 &-8.7\\
21.0 & 3.3 & 0.7 & -1.2 & 2.8 & -7.3 & -8.7 & 505
\end{pmatrix}, \label{eq:fmo2-1}
\end{equation}
with the diagonal offset of 12195 cm$^{-1}$.

The same set of spectral density parameters and temperatures that was used in generation of the FMO-Ib and FMO-II data sets
was used here. LTLME method was used to propagate system's RDM from 0 to 50 ps with 5~fs time-step and three initial states of the electronic
system were considered: sites-1, 6 and 8. The data set contains both diagonal (populations)
and off-diagonal (coherences) elements of the RDM. The calculations was performed with the \textsc{quantum\_HEOM} package\cite{joseph2019quant} with some local modifications to make it compatible for the Hamiltonians with larger dimension. We will refer to this as 
\textsc{modified-quantum\_HEOM} implementation. 

\subsubsection*{FMO-V data set: FMO trimer with local thermalizing Lindblad master equation approach}
Additionally, we also generated a data set for the FMO trimer. The overall excitonic 
Hamiltonian of all three subunits is given by
\begin{equation}
    H_s = \begin{pmatrix}
    H_A & H_B & H_B^T \\
    H_B^T & H_A & H_B \\
    H_B & H_B^T & H_A
    \end{pmatrix}
\end{equation}
where $H_A$ is the subunit Hamiltonian for which we used the same Hamiltonian as in FMO-IV data set (Eq.~\ref{eq:fmo2-1}), while
$H_B$ is the intra-subunit Hamiltonian which is taken from the work of Olbrich \textit{et al.}\cite{olbrich11} and is given by
(in cm$^{-1}$)
\begin{equation}
    H_B = \begin{pmatrix}
    1.0 & 0.3 & -0.6 & 0.7 & 2.3 & 1.5 & 0.9 & 0.1 \\
    1.5 & -0.4 & -2.5 & -1.5 & 7.4 & 5.2 & 1.5 & 0.7 \\
    1.4 & 0.1 & -2.7 & 5.7 & 4.6 & 2.3 & 4.0 & 0.8 \\
    0.3 & 0.5 & 0.7 & 1.9 & -0.6 & -0.4 & 1.9 & -0.8 \\
    0.7 & 0.9 & 1.1 & -0.1 & 1.8 & 0.1 & -0.7 & 1.3 \\
    0.1 & 0.7 & 0.8 & 1.4 & -1.4 & -1.5 & 1.6 & -1.0 \\
    0.3 & 0.2 & -0.7 & 4.8 & -1.6 & 0.1 & 5.7 & -2.3 \\
    0.1 & 0.6 & 1.5 & -1.1 & 4.0 & -3.1 & -5.2 & 3.6
    \end{pmatrix}.
\end{equation}
We propagate dynamics with LTLME from 0 to 50~ps with 5~fs time-step  
for the same parameters as was adopted in calculations for the FMO-Ib---FMO-IV data sets. 
The calculations were 
performed with the \textsc{modified-quantum\_HEOM} implementation 
for the initial excited sites-1, 6 and 8.

\subsubsection*{FMO-VI data set: 8-site FMO model with the hierarchy of equations of motion approach}
The LTLME approach provides only approximate description of  quantum dynamics of the FMO complex.
Therefore, the FMO-I---FMO-V data sets are useful merely for the developing
machine learning models for quantum dynamics studies. For example, they can be used to
train a neural network model which can then be further
improved on more accurate but smaller data sets
(e.g., via transfer learning).
However, LTLME dynamics cannot be used to benchmark
other quantum dynamics methods. In the latter case the high-quality reference data is needed. 

To generate a data set with accurate FMO dynamics we performed HEOM calculations for the
8-site FMO model with the Hamiltonian given by Eq.~\ref{eq:fmo2-1}. 
HEOM calculations were performed using the parallel hierarchy integrator (\textsc{PHI}) 
code.\cite{strumpfer12} The initial data set was chosen on the basis of farthest-point sampling similar
to how it was done
in the FMO-Ib---FMO-V data sets with the only difference being that instead of 500 most distant
sets of parameters that were chosen in the preparation of FMO-Ib---FMO-V data sets, 
1100 most distant set of parameters
were used to prepare the initial FMO-VI data set.
For certain parameters, the RAM requirements exceeded the RAM
of computing nodes available to us (1 TB). Therefore, such parameter sets were excluded from the data set.
Excluded parameters correspond to low temperatures, high reorganization energies, and low cut-off frequency. 
Such strong non-Markovian regimes pose significant challenges in the computational studies of open
quantum systems. 
 Approximately 20\% of the  initial data set was removed because of prohibitive memory requirements.
 We note that even though graphics processing units (GPU) implementations of HEOM (e.g., Ref.~\citenum{kreisbeck11}) 
 are much faster than their CPU-based counterparts, 
they are still limited by the small amount of memory
 in presently available GPUs.

For the remaining 80\% of the data set HEOM calculations were performed for 2.0 ps.
To speed up calculations, an adaptive integration Runge--Kutta--Fehlberg 4/5\cite{fehlberg85} (RKF45) method
was used as implemented in the \textsc{PHI} code. Using adaptive integration reduces both the total computation time 
and memory requirements
but can lead to artifacts if the accuracy threshold is set too large.\cite{strumpfer12} In this work 
the PHI default accuracy threshold of $1\cdot 10^{-6}$ was used. 
The initial integration time step was set to 0.1 fs. In RKF45
the integration time step is varied and, therefore, the output 
comprises time-evolved RDMs on an unevenly spaced time grid.
To obtain the RDMs on an evenly spaced time grid of 0.1 fs, cubic-spline interpolation was used. The interpolation errors were 
examined
on a few cases where 0.1 fs fixed time step integration was feasible. The errors in the
populations were found to be less than
10$^{-5}$ which is much smaller compared to the convergence thresholds discussed below in Technical Validation.
The final FMO-VI data set contains 879 
entries each comprising all the populations and coherences for the
RDM from 0 to 2 ps with the time-step of 0.1 fs.

\section*{Data Records}

All data sets
can be accessed at \url{https://figshare.com/s/ed24594205ab87404238}. The data sets are stored in standard NumPy\cite{oliphant07} binary file format (.npy) 
files.
The following format of file names was adopted in the SB data set
\texttt{2\_epsilon-X\_Delta-1.0\_lambda-Y\_gamma-Z\_beta-XX.npy}, where X denotes the value of the energetic bias ($\tilde{\epsilon}$), Y is the reorganization energy $\tilde{\lambda}$, Z is the cut-off
frequency $\tilde{\gamma}$,
and XX is the inverse temperature $\tilde{\beta}$.
The following format of file names was adopted in all FMO data sets \texttt{X\_initial-Y\_gamma-Z\_lambda-XX\_temp-YY.npy}, where X denotes the number of sites in the FMO model, Y is the initial state, Z is the value of bath frequency, XX is the value of reorganization energy, and YY is the temperature.

\section*{Technical Validation}
\label{sec:val}
Central to the HEOM approach is the assumption that the bath correlation function $C_a(t)$ for site $a$ can be represented 
by an infinite sum of 
exponentially decaying terms $C_a(t) = \sum_k^\infty c_{ak} \exp(-\nu_{ak}t)$, where $\nu_{ak}=2\pi k/\beta \hslash$ are Matsubara 
frequencies. Further, each exponential term leads to a set of auxiliary density matrices
which take into account the non-Markovian evolution of the system's RDM under the influence of bath.
In practice, the summation must be truncated at a finite level, $K$,
which is called Matsubara cut-off and the set of auxiliary density matrices
 needs to be truncated at a finite number $M$. In the truncated set of auxiliary matrices are indexed by
$\mathbf{n}=(n_{10},\ldots,n_{1K},n_{M0},\ldots,n_{MK})$.
 The hierarchy truncation level is given by
 $L=\sum_{a=1}^M\sum_{k=0}^K n_{ak}$, where $n_{ak}$ is the index of an auxiliary density matrix.
 The computational cost of the HEOM method rises steeply with the hierarchy level $L$.\cite{strumpfer12} 
 
The hierarchy truncation level $L$ depends on how non-Markovian the system is. Although, there is some guidance on how to choose the
Matsubara cut-off and hierarchy truncation level based on bath and spectral density parameters,\cite{tanimura20,ishizaki09} in practice, the values of $M$ and $K$ have to be chosen
by requiring the convergence of the RDM to 
acceptable accuracy level. 
In this work HEOM calculations for the SB model were
performed by setting $L=30$ for all temperatures.
The Matsubara cut-off was chosen depending on the temperature as follows: for $\beta$ = 0.1  $K$ was set to 2;
for $\beta = 0.25, K = 3$, for $\beta = 0.5, K = 3$, for $\beta = 0.75, K = 4$, and for $\beta = 1.0, K = 5$.
 These values are chosen sufficiently high to ensure the convergence of the populations with respect to
 $K$ and $L$. Choosing high truncation levels in the HEOM
 calculations of a TLS does not present a problem given the presently 
 available computational resources. 
 
 Similar approach of taking excessively large values of $K$ and $L$, however, is infeasible in the FMO calculations
 because the computational cost of HEOM grows steeply with the size of the quantum system. Therefore, the following 
 approach was adopted for the HEOM calculations of the 8-site FMO model (FMO-VI data set). Starting from
 $K=0$ and $L=1$, $K$ was increased until the maximum difference in the populations between calculations with $K$ and $K+1$
 falls below a threshold $\Delta$, i.e.,
 \begin{equation}
     \delta = \max_{\substack{n=1,\ldots,N_{el} \\ t=0,\ldots,t_{max}}} \bigg| \rho_{n,n}^{K,L}(t) - \rho_{n,n}^{K+1,L}(t) \bigg| < \Delta. \label{eq:cvg}
 \end{equation}
 When Eq.\ref{eq:cvg} is satisfied for a given $\Delta$ the convergence with respect to Matsubara cut-off is deemed
 to have been achieved.
 Then, for a fixed $K$ a
 series of calculations were performed with increasing values of $L$ until the maximum difference in populations 
 between two consecutive calculations becomes less than the same threshold value $\Delta$. When this condition is satisfied 
 the convergence with respect to hierarchy truncation level as well as the overall convergence is declared.
 These steps were performed in the HEOM
 calculations for each parameter set for an 8-site FMO model until either the overall convergence is achieved or $K$ and/or
 $L$ become large enough so the calculation becomes intractable exceeding RAM available on our machines (1~TB). 
 
\begin{figure}[ht]
\centering
\includegraphics[width=\linewidth]{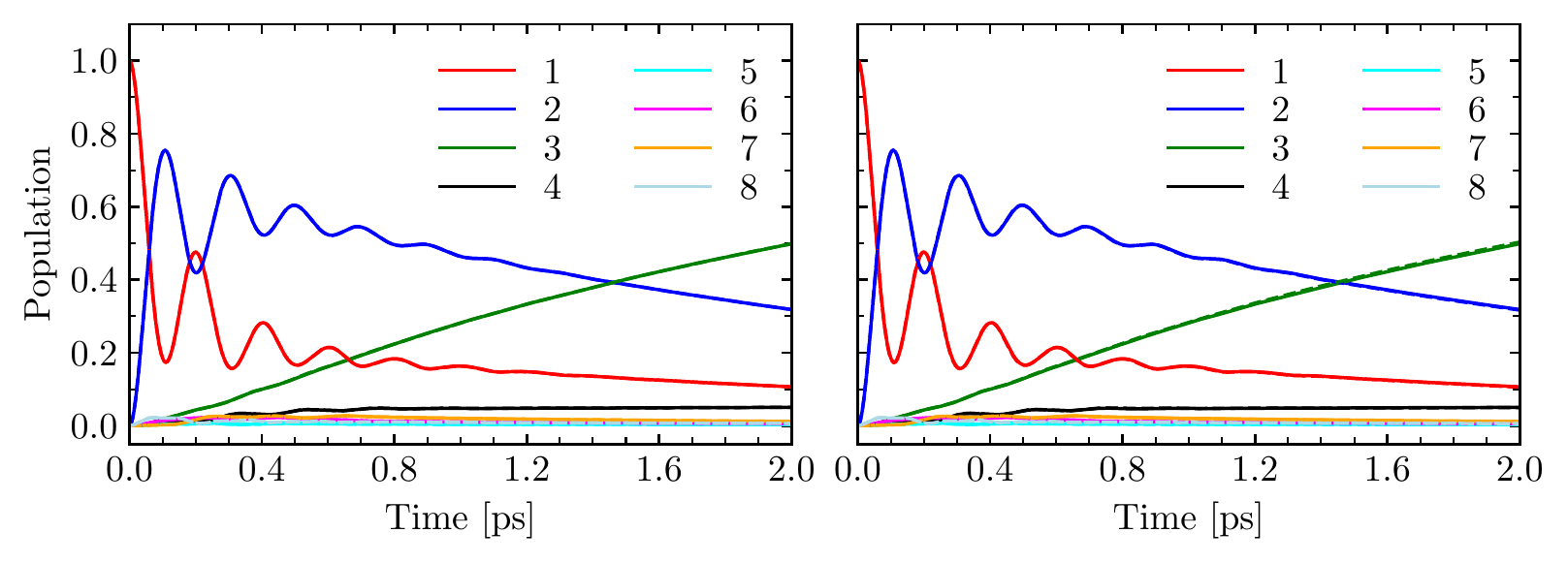}
\caption{Example of the technical validation of the convergence of HEOM calculations for the following parameters
$T=30$ K, $\lambda=70$ cm$^{-1}$, $\gamma=500$ cm$^{-1}$. The convergence within $\Delta=0.01$ threshold was achieved
for $K=7$ and $L=4$. Left plot shows the populations obtained with $K=7$ and $L=4$ (solid lines) compared to populations obtained
with $K=7$ and $L=5$ (dashed lines) for all 8 sites. The largest population difference is $\delta=3.14\cdot 10^{-4}$.
The right plot shows the populations obtained with $K=7$ and $L=4$ (solid lines) compared to the populations obtained
with $K=8$ and $L=4$. The largest population difference is $\delta=4.62\cdot 10^{-3}$. In both cases the difference
is very small illustrating the validity of the chosen threshold.}
\label{fig:cvg}
\end{figure}

\begin{figure}[ht]
\centering
\includegraphics[width=\linewidth]{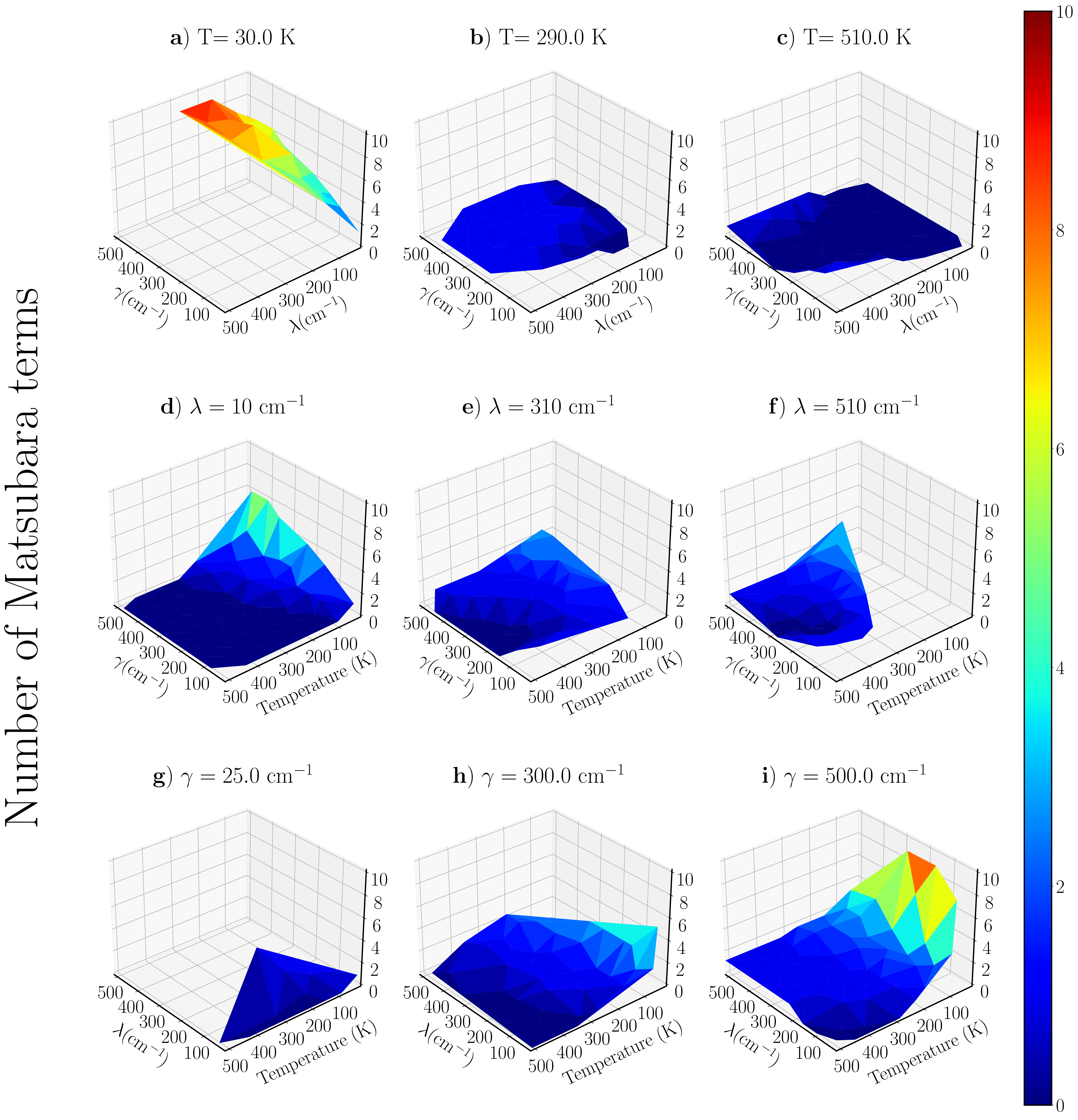}
\caption{Number of Matsubara terms (Matsubara cut-off, $M$) required for converging populations of an
8-site FMO model with the system Hamiltonian given by Eq.~\ref{eq:fmo2-1} for three selected temperatures ($T$),
reorganization energies ($\lambda$), and bath cut-off frequencies ($\gamma$).}
\label{fig:matsubara}
\end{figure}

\begin{figure}[ht]
\centering
\includegraphics[width=\linewidth]{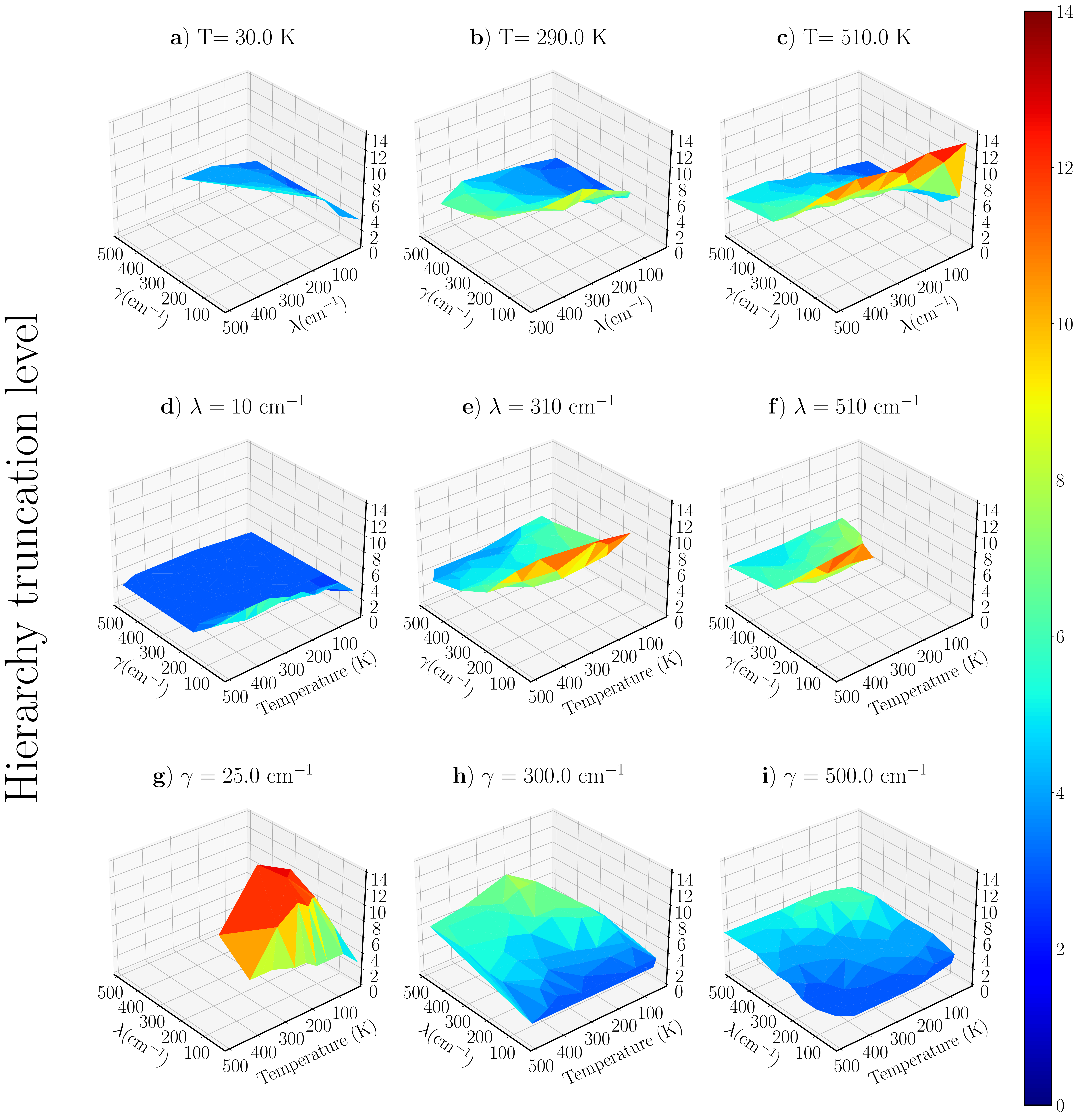}
\caption{Hierarchy truncation level $L$ required for converging populations of an
8-site FMO model with the system Hamiltonian given by Eq.~\ref{eq:fmo2-1} for three selected temperatures ($T$),
reorganization energies ($\lambda$), and bath cut-off frequencies ($\gamma$).}
\label{fig:hierarchy}
\end{figure}

In this work we set the threshold $\Delta = 0.01$. This threshold was chosen such that the population errors would be 
 almost imperceptible which is illustrated in Figure~\ref{fig:cvg}. This data set is converged enough to be helpful in
 benchmarks of approximate methods describing quantum dynamics because the errors of these methods often exceed the threshold
 used in this work. 
 Additionally, Figures~\ref{fig:matsubara} and ~\ref{fig:hierarchy} show  the number of Matsubara terms and the hierarchy truncation level  required
 for achieving the overall convergence depending on spectral density parameters and temperature.  

\section*{Usage Notes}
A Python package for extracting data is provided together with the data set and can be accessed at \hyperlink{https://github.com/Arif-PhyChem/QD3SET}{https://github.com/Arif-PhyChem/QD3SET}.

\section*{Code availability}
\textsc{PHI} code (version 1.0) used in HEOM calculations was downloaded from \href{http://www.ks.uiuc.edu/Research/phi/}{http://www.ks.uiuc.edu/Research/phi/}.
\textsc{QuTip} software package (version 4.6) used in HEOM calculations of the spin-boson model and  was downloaded from \href{https://qutip.org/}{https://qutip.org/}. LTLME calculations of FMO models were performed with the basic \textsc{quantum\_HEOM} package \href{https://github.com/jwa7/quantum_HEOM}{https://github.com/jwa7/quantum\_HEOM} and was modified 
to enable compatibility with the Hamiltonian of larger than the default dimension.

\bibliography{main}


\section*{Acknowledgements} 
A.A.K. acknowledges the Ralph E. Powe Junior Faculty Enhancement Award from Oak Ridge Associated Universities. 
This work was also supported by General University Research (GUR) Grants and 
 startup funds of the College of Arts and Sciences and the Department of Physics and Astronomy of the University of Delaware. 
P.O.D. acknowledges funding by the National Natural Science Foundation of China (No. 22003051 and funding via the Outstanding Youth Scholars (Overseas, 2021) project), the Fundamental Research Funds for the Central Universities (No. 20720210092), and via the Lab project of the State Key Laboratory of Physical Chemistry of Solid Surfaces.
This research was supported in part through the use of Data Science Institute (DSI) computational resources at the University of Delaware.
 Calculations were also
performed with high-performance computing resources provided by the Xiamen University.

\section*{Author contributions statement}
P.O.D. and A.U. conceived the idea of creating a HEOM-based spin-boson database. 
A.U. conceived the idea of creating LTLME-based database for FMO complex. A.A.K conceived the idea of creating an FMO dataset
with the HEOM method.
A.U. performed the HEOM calculations for spin-boson along with the LTLME calculations for FMO complex. 
A.U. wrote the provided package for easy extraction of the data. 
A.A.K and L.E.H.R. performed the calculations and created database files for the FMO-VI data set.
All authors analysed the results. A.A.K. took the lead in writing the original draft of the manuscript. 
All authors reviewed and approved the manuscript. 

\section*{Competing interests} 
The authors declare no competing interest.

\section*{Figures \& Tables}




\begin{table}[ht]
\centering
\resizebox{\textwidth}{!}{\begin{tabular}{|l|l|l|l|l|l|l|l|l|l|}
\hline
\makecell{Data set} & System & Hamiltonian(s) & Method & \makecell{Data set \\ size} & Cases 
 & \makecell{Propagation \\ time\\  (time-step)} & Package & Parameter space & References\\
\hline
SB & SB & SB & HEOM & \multirow{4}{*}{1000} & \makecell{Symmetric \\ and \\ Asymmetric} & 20$/\Delta$ (0.05$/\Delta$) & \textsc{QuTiP}\cite{johansson12} & $\mathcal{E}^b$ & \makecell{Regenerated \\ based on Ref.~\citenum{ullah2021speeding}}\\
\cline{1-4}\cline{6-10}
FMO-Ia & \multirow{3}{*}{7-site FMO} & \multirow{2}{*}{\makecell{Adolphs and \\ Renger\cite{adolphs06}}} & \multirow{6}{*}{LTLME} &  & \multirow{3}{*}{Sites 1 and 6}  & 1~ns (5~fs) & 
\textsc{quantum\_HEOM}\cite{joseph2019quant,joseph2020thesis} &  $\mathcal{F}^c$ & \makecell{Regenerated \\ based on Ref.~\citenum{ullah2022predicting}} \\
\cline{1-1} \cline{7-10} 
FMO-Ib &  & &  &  &  &  \multirow{6}{*}{50~ps (5~fs)} &  \multirow{5}{*}{\makecell{\textsc{modified-}\\\textsc{quantum\_HEOM}}$^a$} & \multirow{5}{*}{\makecell{$\mathcal{G}^d$}} &\multirow{6}{*}{This work}  \\
\cline{1-1} \cline{3-3}
FMO-II &  & Cho\cite{cho05} &  &  &  &  &  &  &\\
\cline{1-3} \cline{5-6}
FMO-III & \multirow{2}{*}{8-site FMO} & Jia\cite{jia15} &  &  &  &  &  &  &\\
\cline{1-1} \cline{3-3}
FMO-IV &   & \multirow{3}{*}{\makecell{Busch\cite{busch11} \\ and \\ Olbrich\cite{olbrich11}}} &  & 1500 & \makecell{Sites 1, 6 \\ and 8} &  &  &  &\\
\cline{1-2}
FMO-V & FMO trimer &  &  &  &  &  &  &  & \\
\cline{1-2}\cline{4-9}
FMO-VI & 8-site FMO &  & HEOM & 879 & Site 1 & 2~ps (0.1~fs) & \textsc{PHI}\cite{strumpfer12} & $\mathcal{H}^e$ &\\
\hline
\end{tabular}}
\caption{\label{tab:tab1} Summary of all data sets. See more details in the main text. Here ``SB" stands for spin-boson model. $^a$\textsc{modified-quantum\_HEOM} is the \textsc{quantum\_HEOM} package with some local modifications to make it compatible for larger Hamiltonians. $^b$In parameter space $\mathcal{E}$, we define $\tilde{\epsilon}=\epsilon/\Delta=\{0, 1\}$, $\tilde{\lambda}=\lambda/\Delta=\{0.1$, $0.2$, $0.3$, $0.4$, $0.5$, $0.6$, $0.7$, $0.8$, $0.9$, $1.0\}$, 
$\tilde{\gamma}=\gamma/\Delta =\{1$, $2$, $3$, $4$, $5$, $6$, $7$, $8$, $9$, $10\}$, and $\tilde{\beta}=\beta \Delta=\{0.1$, $0.25$, $0.5$, $0.75$, $1\}$,
where the tunneling matrix element $\Delta$ is set as an energy unit. For all combinations of these
parameters the system's RDM was propagated. $^c$In parameter space $\mathcal{F}$, we choose the top 500 (most distant)
combinations of ($\lambda$ , $\gamma$, $T$) based on farthest-point sampling. Parameter range for each dimension is $\lambda$ = $\{$10, 40, 70, \ldots, 310$\}$ cm$^{-1}$,
$\gamma$ = $\{$25, 50, 75, \ldots, 300$\}$ fs rad$^{-1}$,
and T = $\{$30, 50, 70, \ldots, 310$\}$ K. $^d$In parameter space $\mathcal{G}$, we adopt the same procedure as in parameter space $\mathcal{F}$ and choose the most distant 500 points (based on farthest-point sampling) from 3D space ($\lambda$ , $\gamma$, $T$) where $\lambda$ = $\{$10, 40, 70, \ldots, 520$\}$ cm$^{-1}$,
$\gamma$ = $\{$25, 50, 75, \ldots, 500$\}$ cm$^{-1}$,
and T = $\{$30, 50, 70, \ldots, 510$\}$ K. $^e$In parameter space $\mathcal{H}$, parameters range remains the same as  in $\mathcal{G}$. In addition, the same farthest point sampling was adopted but with the only difference being that instead of 500, 1100 most distant set of parameters
were chosen. Approximately 20\% of the  initial data set was removed because of the prohibitive memory requirements. For the remaining 80\% of the data set, HEOM calculations were performed for 2.0~ps using 0.1~fs as a time step. }
\end{table}

\end{document}